\documentclass[aps,prx,twocolumn,superscriptaddress,showpacs,floatfix]{revtex4}

\usepackage{graphicx,color}
\usepackage{dcolumn}
\usepackage{bm}
\usepackage{stmaryrd}
\usepackage{latexsym}
\usepackage{amssymb}
\usepackage{amsfonts}
\usepackage{amsmath}

\begin{document}


\title{The effects of electron correlation and spin-orbit coupling in the isovalent Pd-doped superconductor SrPt$_3$P}

\author{Kangkang Hu}\affiliation{State Key Laboratory of
Functional Materials for Informatics and Shanghai Center for
Superconductivity, Shanghai Institute of Microsystem and Information
Technology, Chinese Academy of Sciences, Shanghai 200050,
China}\affiliation{Shanghai Key Laboratory of High Temperature
Superconductors, Shanghai University, Shanghai 200444, China}

\author{Bo Gao}
\affiliation{State Key Laboratory of Functional Materials for
Informatics and Shanghai Center for Superconductivity, Shanghai
Institute of Microsystem and Information Technology, Chinese Academy
of Sciences, Shanghai 200050, China}

\author{Qiucheng Ji}
\affiliation{State Key Laboratory of Functional Materials for Informatics
and Shanghai Center for Superconductivity, Shanghai Institute of Microsystem
and Information Technology, Chinese Academy of Sciences, Shanghai 200050, China}

\author{Yonghui Ma}
 \affiliation{State Key
Laboratory of Functional Materials for Informatics and Shanghai
Center for Superconductivity, Shanghai Institute of Microsystem and
Information Technology, Chinese Academy of Sciences, Shanghai
200050, China}\affiliation{School of Physical Science and
Technology, ShanghaiTech University, Shanghai 201210, China}

\author{Wei Li}
\email[]{liwei@mail.sim.ac.cn}\affiliation{State Key Laboratory of
Functional Materials for Informatics and Shanghai Center for
Superconductivity, Shanghai Institute of Microsystem and Information
Technology, Chinese Academy of Sciences, Shanghai 200050, China}
\affiliation{CAS-Shanghai Science Research Center, Shanghai 201203, China}

\author{Xuguang Xu}
\affiliation{School of Physical Science and
Technology, ShanghaiTech University, Shanghai 201210, China}

\author{Hui Zhang}
\affiliation{CAS Key Laboratory of Materials for Energy Conversion,
Shanghai Institute of Ceramics, Chinese Academy of Sciences,
Shanghai 200050, China}

\author{Gang Mu}
\email[]{mugang@mail.sim.ac.cn} \affiliation{State Key Laboratory of
Functional Materials for Informatics and Shanghai Center for
Superconductivity, Shanghai Institute of Microsystem and Information
Technology, Chinese Academy of Sciences, Shanghai 200050, China}

\author{Fuqiang Huang}
\affiliation{CAS Key Laboratory of Materials for Energy Conversion,
Shanghai Institute of Ceramics, Chinese Academy of Sciences,
Shanghai 200050, China}

\author{Chuanbing Cai}\affiliation{Shanghai Key
Laboratory of High Temperature Superconductors, Shanghai University,
Shanghai 200444, China}

\author{Xiaoming Xie}
\affiliation{State Key Laboratory of Functional Materials for Informatics and
Shanghai Center for Superconductivity, Shanghai Institute of Microsystem and
Information Technology, Chinese Academy of Sciences, Shanghai 200050, China}

\author{Mianheng Jiang}
\affiliation{State Key
Laboratory of Functional Materials for Informatics and Shanghai
Center for Superconductivity, Shanghai Institute of Microsystem and
Information Technology, Chinese Academy of Sciences, Shanghai
200050, China}\affiliation{School of Physical Science and
Technology, ShanghaiTech University, Shanghai 201210, China}

\begin{abstract}
We present a systematical study on the roles of electron correlation
and spin-orbit coupling in the isovalent Pd-doped superconductor
SrPt$_3$P. By using solid state reaction method, we fabricated the
strong spin-orbit coupling superconductors Sr(Pt$_{1-x}$Pd$_x$)$_3$P
with strong electron correlated Pd dopant of the $4d$ orbital. As
increasing the isovalent Pd concentrations without introducing any
extra electron/hole carriers, the superconducting transition
temperature $T_c$ decreases monotonously, which suggests the
existence of competition between spin-orbit coupling and electron
correlation in the superconducting state. In addition, the
electronic band structure calculations demonstrate that the strength
of electron susceptibility is suppressed gradually by the Pd dopant
suggesting the incompatible relation between spin-orbit coupling and
electron correlation, which is also consistent with experimental
measurements. Our results provide significant insights in the
natures of the interplay between the spin-orbit coupling and the
electron correlation in superconductivity, and may pave a way for
understanding the mechanism of superconductivity in this
5d-metal-based compound.
\end{abstract}

\date{\today}
\maketitle

\section{Introduction}
The study of strong interplay among charge, spin, orbital, and
lattice degrees of freedom in transition metal compounds has
triggered enormous research interests in the communities of
condensed matter physics and material physics. One of the most
prominent example is the unconventional high-transition temperature
(high-$T_c$) superconductivity induced by the strong electron
correlation in the copper oxide and iron-based
superconductors~\cite{cuprates,pnictides}. In those materials,
apparently, the spin degree of freedom plays a vital rule and the
orbital degree of freedom is decoupled from that of spin. However,
in strong spin-orbit coupling superconductors, such as the recent
discovered platinum-based superconductors APt$_3$P (A = Ca, Sr and
La)~\cite{3,5,7,9,11,Si-doping,IXS,NMR}, the systems display weak
electron correlation effect. The relation between strong electron
correlation and spin-orbit coupling still remains unclear. To
clarify the nature of the interplay between spin-orbit coupling and
electron correlation in superconductors is a crucial issue not only
in condensed matter physics, but also in material science. Motivated
by this issue, we focus our great attentions on the isovalent Pd
doped platinum-based superconductors SrPt$_3$P, where the strength
of the spin-orbit coupling and electron correlation can be tuned by
the Pd concentration without introducing any extra electron or hole
carriers into the system. This physics is quite different from the
case that of the Lanthanum-based superconductor LaPt$_3$P, where the
extra electron is injected into the system as strontium is replaced
by Lanthanum leading to the decrease of the total density of states
(DOS) at the Fermi level, which results in the suppression of
superconducting transition temperature $T_c$~\cite{11}.

In this paper, the Pd dopant with strong electron correlation of
$4d$ orbital was successfully substituted to the site of Pt in
strong spin-orbit coupling superconductor SrPt$_3$P, which was
confirmed by the x-ray diffraction measurements. The crystal lattice
was found to shrink along the c-axis and expand along the a-axis
monotonously as increasing the Pd concentrations. Importantly, we
find that the superconducting transition temperature $T_c$ decreases
with Pd doping, which could not be attributed to the physics of
changes of the DOS at the Fermi level and the impurity scattering.
Thus, it suggests that the system exhibits a competition between
electron correlation and spin-orbit coupling in our system. Such a
competing relation is also consistent with the first-principles
calculation, which show that the strength of electron susceptibility
is suppressed gradually as increasing Pd dopants.

\section{Experiments}

The samples in this work were prepared via solid state reaction from
the pure elements.~\cite{MgIrB} Firstly, we put stoichiometric
amounts of platinum powder (purity 99.97\%, Alfa Aesar), red
phosphorus powder (purity 99.9\%, Aladdin), and strontium pieces
(purity 98\%+, Alfa Aesar) together and ground them in a mortar.
After that, the mixture was pressed into a small pellet and then
sealed in a clean vacuum quartz tube. All the weighing and mixing
procedures were carried out in a glove box with a protective argon
atmosphere. The tube was heated up to 400 $^\circ$C and held for 10
hours to prevent red phosphorus from volatilizing so quickly, and
calcined at 900 $^\circ$C for 2 days. The sintered pellet was
reground and further annealed at 900 $^\circ$C within an
argon-filled quartz tubes for 3 days. The doped samples
Sr(Pt$_{1-x}$Pd$_x$)$_3$P were prepared with adding corresponding
amount of palladium powder (purity 99.95\%, Alfa Aesar) using the
same method as mentioned above.

The structure of the obtained samples were checked using a DX-2700
type powder x-ray diffractometer. The magnetic susceptibility
measurements were carried out on the magnetic property measurement
system (Quantum Design, MPMS 3). The electrical resistance was
measured using a four-probe technique on the physical property
measurement system (Quantum Design, PPMS).

\section{Results and Discussions}

\subsection{Crystal Structure}

\begin{figure}
\includegraphics[width=10cm]{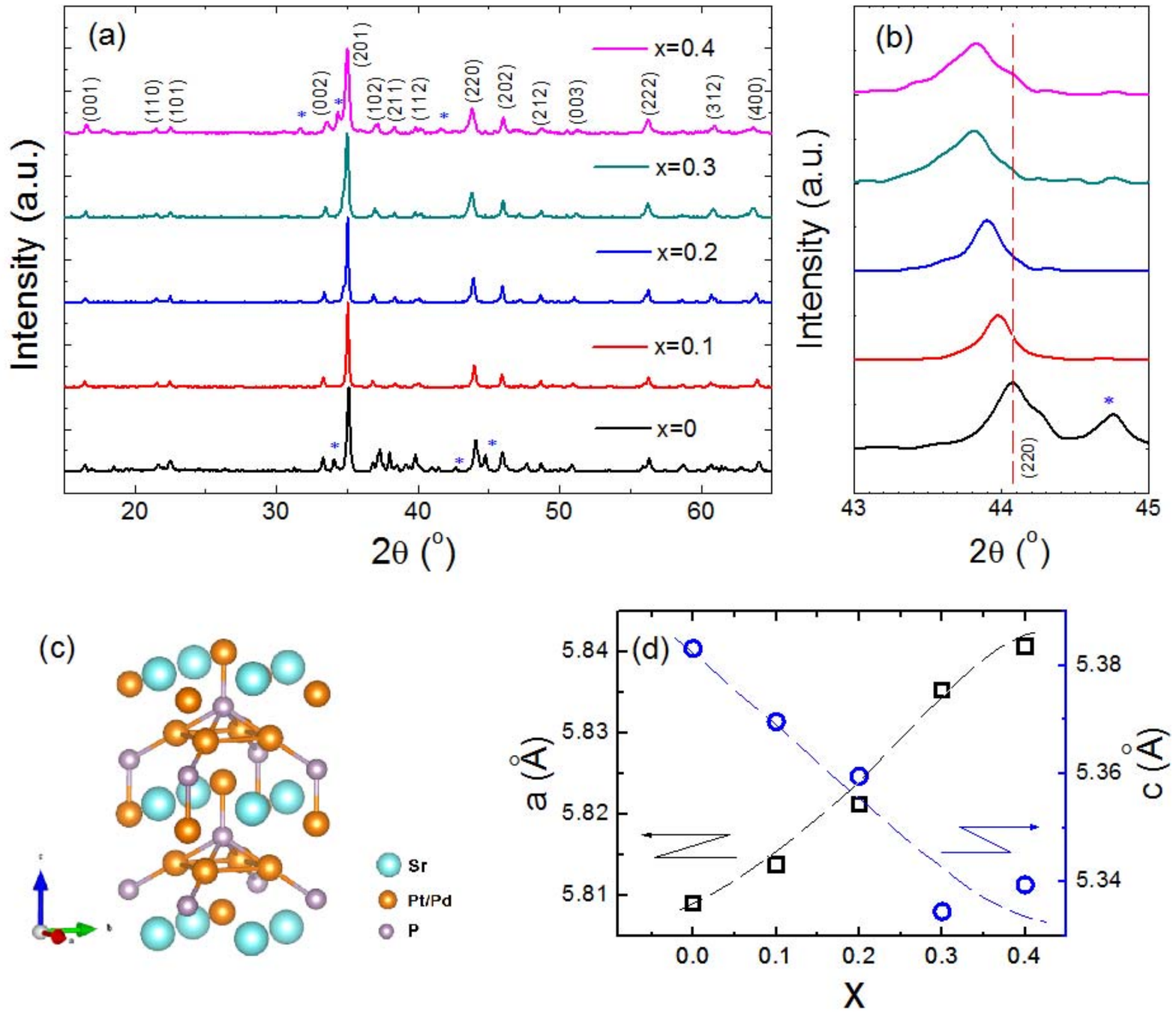}
\centering\caption {(color online) (a) X-ray diffraction patterns
for the Sr(Pt$_{1-x}$Pd$_x$)$_3$P samples with $0 \leq x \leq 0.4$.
Small amount of impurity phases are indexed by the blue asterisks.
(b) An enlarged view of the (220) peak. It is clear that the peak
shifts monotonously to the left direction with doping. (c) Crystal
structure of SrPt$_3$P. Pd element was substituted into the site of
Pt in this work. (d) Doping dependence of the lattice constants
along a-axis and c-axis.} \label{fig1}
\end{figure}

The crystal structure of this system has been reported by T.
Takayama et al. previously~\cite{3}. Here we concentrate to the
influences on the crystal lattice by the Pd substitution. For this
reason, we measured the powder X-ray diffraction (XRD) on this
series of materials and the diffraction patterns are shown in Fig. 1
(a). It is found that the main diffraction peaks can be indexed to
the tetragonal structure with the space group $P4/nmm$ as shown in
Fig. 1(c). The black line represents the parent compound SrPt$_3$P.
The asterisks refer to some unknown impurities which was also
observed by T. Takayama et al.~\cite{3} With the increase of the
amount of Pd ($x$), there is no obvious increase of the impurity
phases. However, the positions of diffraction peaks start to move
apparently when $x$ changes from 0 to 0.4, which points to the
variation of the size of the crystal lattice. This tendency can be
seen clearly in Fig. 1(b), where we enlarge the region near the
(220) diffraction peak as an example.

To investigate the influences on the crystal lattice by the Pd
substitution quantitatively, we obtained the lattice parameters by
fitting the XRD data using the software Powder-X.~\cite{Powder-X}
The results are shown in Fig. 1(d). As is shown in the graph, the
crystal lattice shows a shrinkage along the c-axis, while it expands
along the a-axis with the increase of doping. This tendency is very
similar to that observed in 4d- and 5d-metal-doped iron arsenides
SrFe$_{2-x}$M$_x$As$_2$ (M=Rh, Ir,Pd),~\cite{4d5d} which proves that
Pd atoms take the place of Pt atoms leading to the alteration of
cell parameters. In the high doping region above 0.4, the variation
of the lattice parameters becomes gentle, indicating the limit of
the chemical substitution. We note that this is a common phenomenon
in some chemical doped materials and such a saturated features has
been reported elsewhere.~\cite{BaTiAsO}

\subsection{Superconducting Properties}

\begin{figure}
\includegraphics[width=9cm]{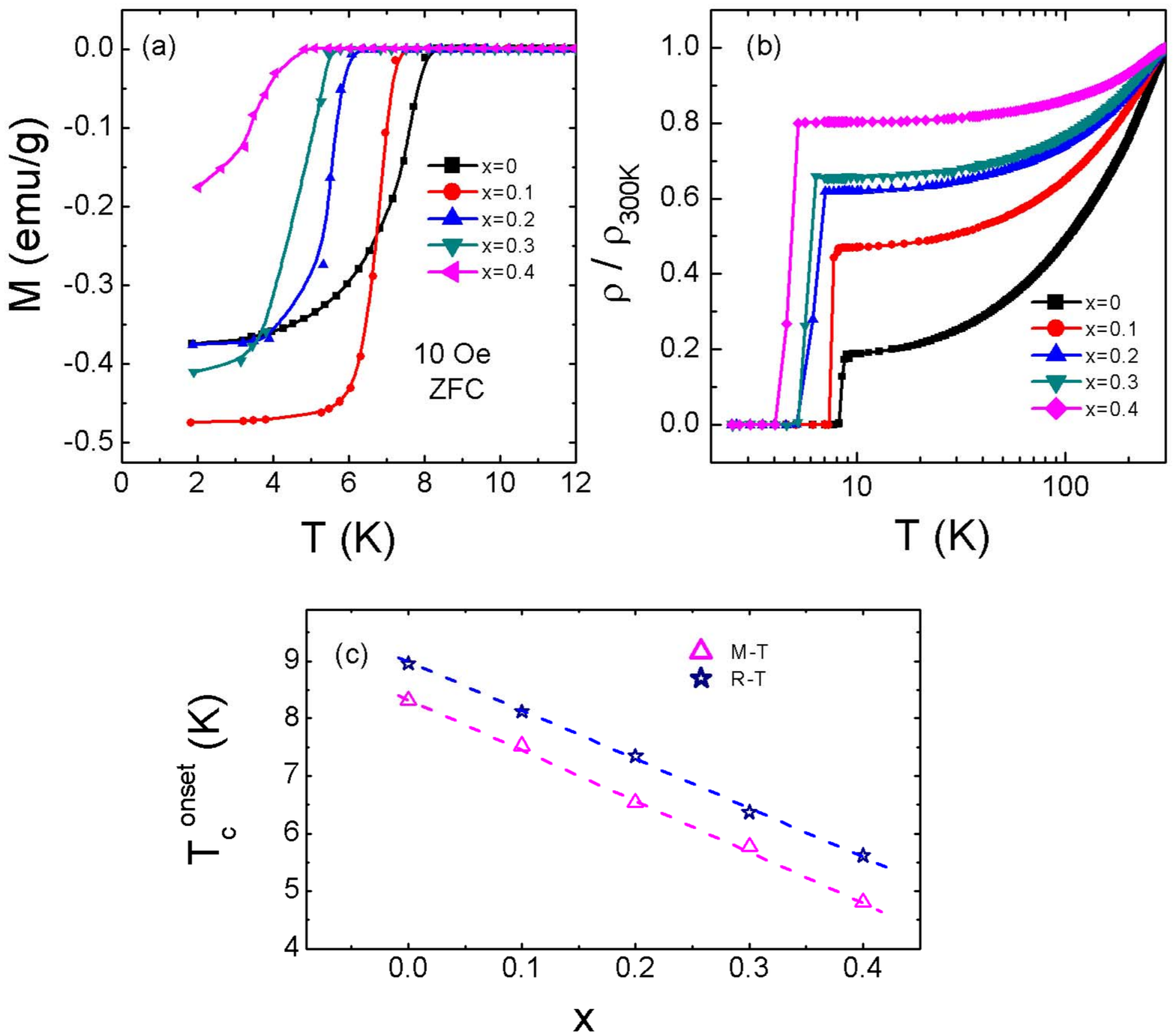}
\centering\caption {(color online) (a) Temperature dependence of the
dc magnetic susceptibility for Sr(Pt$_{1-x}$Pd$_x$)$_3$P samples
with $0 \leq x \leq 0.4$. The data were measured via the
zero-field-cooling (ZFC) mode under the field of 10 Oe. (b)
Temperature dependence of the normalized resistivity. (c) Doping
dependence of the onset superconducting transition temperature. The
blue and purple lines were defined by the magnetization measurements
and the resistivity measurements respectively.} \label{fig2}
\end{figure}

In order to study the effect of Pd-doping on the superconducting
properties, we performed the temperature dependent magnetization and
resistivity measurements on all the samples we synthesized. The
temperature dependence of dc magnetic susceptibility is shown in
Fig. 2(a), where we can see a sharp decline to the negative sides of
the data for each sample with different doping. The clear
diamagnetic signal indicates the occurrence of the superconducting
transition and the onset transition point defines the critical
transition temperature $T_c$. With the increase of amount of Pd, the
value of $T_c$ reduces apparently. This tendency is confirmed by the
resistivity data, which is shown in Fig. 2(b). Clear superconducting
transitions to zero resistance were observed on all the samples with
different doping levels. The onset transition temperatures
determined from this figure, along with that determined from the
$M-T$ curves, are displayed in Fig. 2(c). The blue symbols were
obtained from the magnetization measurements while the purple ones
were defined by the resistivity measurements. It can be seen that
the two sets of $T_c$ values evolute parallelly and decrease
linearly with doping.

\begin{figure*}
\includegraphics[width=15cm]{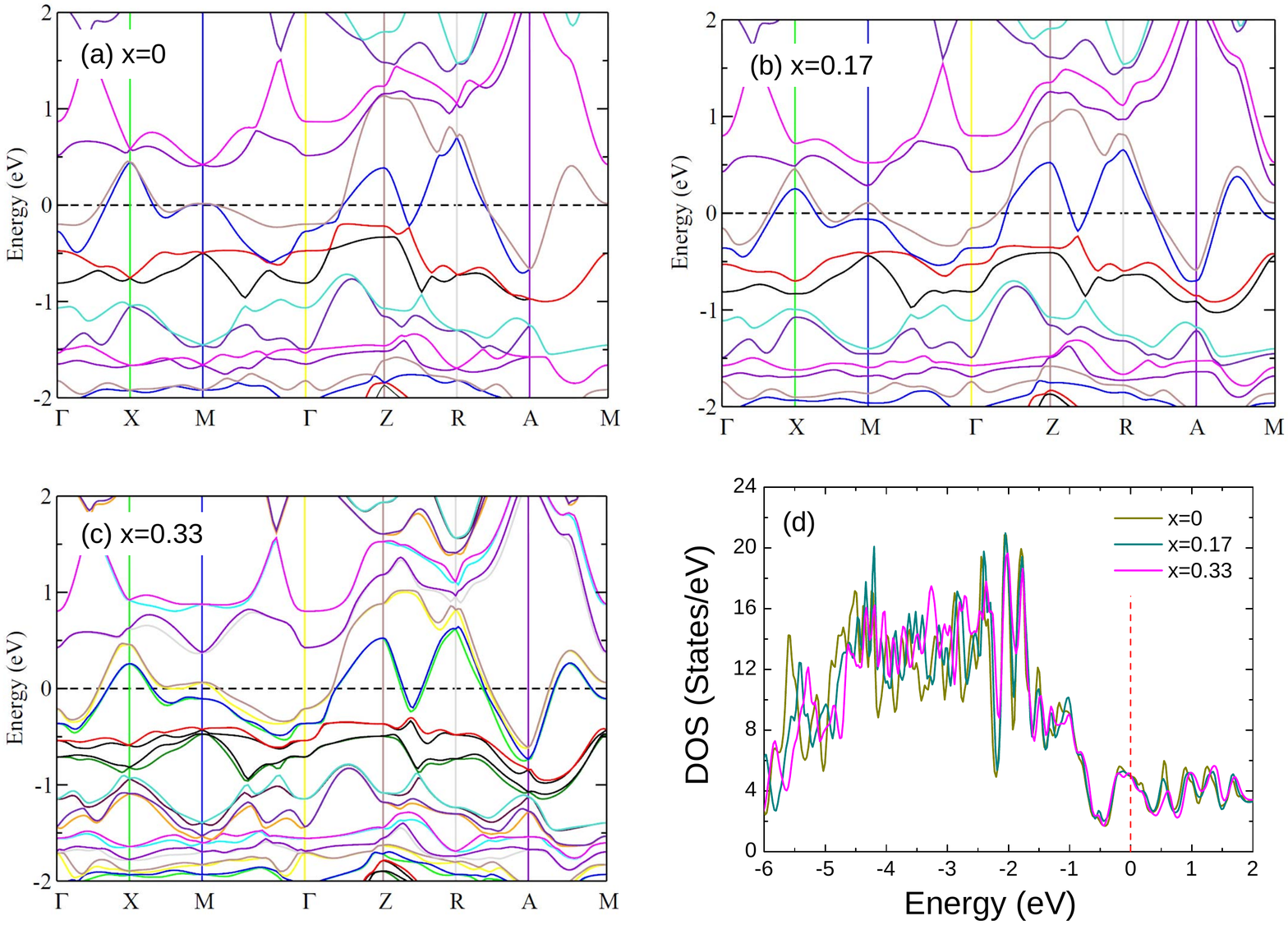}
\centering\caption {(color online) The electronic band structure
calculations for Sr(Pt$_{1-x}$Pd$_x$)$_3$P with doping (a) $x$ = 0,
(b) $x$ = 0.17, and (c) $x$ = 0.33. (d) The corresponding density of
states (DOS) to the three samples. The Fermi energy was set to zero
(dashed lines).} \label{fig3}
\end{figure*}

Generally speaking, the suppression of superconducting transition
temperature $T_c$ is usually related to the changes of DOS at Fermi
level, as has been reported in the Lanthanum replaced superconductor
LaPt$_3$P by comparing with SrPt$_3$P~\cite{11}. However, the
calculated DOS by means of density functional theory demonstrated
that the DOS around the Fermi level remains almost the same for
different concentrations of Pd dopants [see Fig.~\ref{fig3}(d)],
which will be discussed in the next section in detail. This result
is reasonable since the Palladium is isovalent to Platinum making
the carriers of the system remain unchanged. Thus, it rules out the
possibility of the effect from the changes of DOS on
superconductivity in our system.

Another issue is the possible impurity scattering effect~\cite{HKK},
which comes from the Palladium  substitution for Platinum. This
possibility can also be ruled out since SrPt$_3$P was reported to be
an s-wave superconductor.~\cite{3,NMR} According to the Anderson's
theorem~\cite{Anderson}, the non-magnetic impurity does not lead to
an apparent pair-breaking effect in a conventional s-wave
superconductor, and thus does not suppress the transition
temperature $T_c$ apparently, which is in sharp contrast to that in
a d-wave superconductor, where the gap function has a nodal line and
the zero energy excitation spectra can be modified significantly by
non-magnetic impurities leading to the suppression of
$T_c$~\cite{Chien}.

Since the experimental observation of the suppression of the $T_c$
does not originate from the changes of DOS at the Fermi level and
the impurity scattering effect, the $T_c$ suppression is likely to
be related to the interplay between the strong spin-orbit coupling
and the electron correlation. As we have known, the spin-orbit
coupling strength is proportional to $Z^4$ (where $Z$ is the atomic
number; $Z_{Pt}=78$ for Pt and $Z_{Pd}=46$ for Pd)~\cite{WLi}, the
ratio $\gamma=(\frac{Z_{Pt}}{Z_{Pd}})^4=8.3$ and consequently the
spin-orbit coupling strength will be decreased dramatically when the
Platinum is substituted by Palladium. In addition, the bandwidth of
Pd $4d$ orbitals is narrower than that of Pt $5d$ orbitals making
the electron correlation in Pd $4d$ orbital electrons is stronger
than that in Pt $5d$ orbital electrons. Therefore, when the
enhancement of the strength of electron correlation by Palladium
dopants is comparable to that of spin-orbit coupling, the interplay
between electron correlation and the spin-orbit coupling will play a
crucial rule in superconductivity and affect the superconducting
transition temperature $T_c$.

\subsection{First-Principles Calculations}

\begin{figure}
\includegraphics[width=9cm]{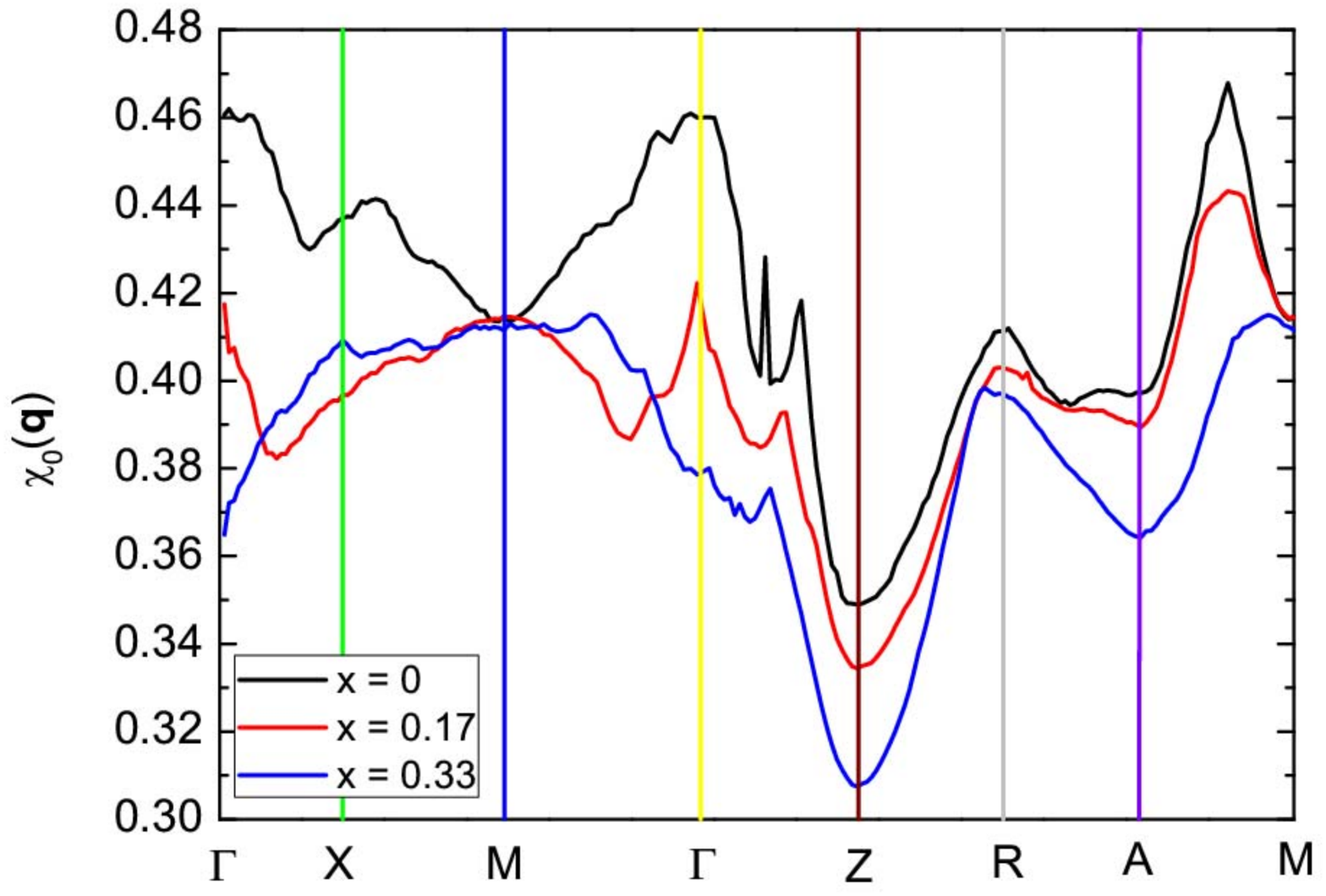}
\centering\caption {(color online) The real part of bare electron
susceptibility $\chi_0(\vec{q})$ along the path between high
symmetric momentum points.} \label{fig4}
\end{figure}

In order to clarify the interplay between electron correlation and
spin-orbit coupling and the origins for the suppression of $T_c$ by
Pd doping in superconductor Sr(Pt$_{1-x}$Pd$_x$)$_3$P
theoretically, we carried out the first-principles band structure
calculations on the Pd dopant dependent samples with $x$ = 0, $x$ =
0.17, and $x$ = 0.33. The calculations were performed by using the
pseudopotential-based code VASP~\cite{VASP} within the
Perdew-Burke-Ernzerhof~\cite{PBE} generalized gradient
approximation. Throughout the theoretical calculations, a $500$ eV
cutoff in the plane wave expansion and a $12\times 12\times 12$
Monkhorst $\vec{k}$ grid were chosen to ensure the calculation with
an accuracy of $10^{-5}$ eV. The atomic coordinates were obtained by
the relaxation based on the lattice parameters from experiments.
Additionally, the spin-orbit coupling had been included throughout
the calculations.

In Fig. 3, we show the energy band structures and their
corresponding DOS. By comparing the energy band structures and DOS
with different Pd dopants, the low energy features of electronic
structures remain almost the same, except for lifting the degenerate
bands and changing their dispersion at some high symmetric momentum
points around the Fermi level. This originates from the nature of
dopant Pd. When the atom Pt in SrPt$_3$P is partially substituted by
Pd, the high symmetry of point group of system was lowered leading
to the lifting of some degenerate energy bands. In addition, since
the Pd atom with $4d$ orbital has a stronger electron correlation
than that of Pt with $5d$ orbital, the strong electron-electron
interaction changes the low energy dispersion and makes it more flat
with a large effective mass.

To quantitatively find the relation between electron correlation and
spin-orbit coupling in the present isovalent Pd doped system, we
further carried out the electron susceptibility calculations. The
bare electron susceptibility, $\chi_0(\vec{q})$, is given by:
\begin{eqnarray*}
\chi_0(\vec{q})=\frac{1}{N_{\mathbf{k}}}\sum_{\mu\nu\mathbf{k}}\frac{|\langle\mathbf{k}+\vec{q},\mu|\mathbf{k},\nu\rangle|^2}
{E_{\mu,\mathbf{k}+\vec{q}}-E_{\nu,\mathbf{k}}+i0^+}[f(E_{\nu,\mathbf{k}})-f(E_{\mu,\mathbf{k}+\vec{q}})],
\end{eqnarray*}
where $E_{\mu,\mathbf{k}}$ represents the band energy measured at
Fermi level $E_F$, and $f(E_{\mu,\mathbf{k}})$ is the Fermi-Dirac
distribution function for an eigenstate, $|\mathbf{k},\mu\rangle$.
In addition, $N_{\mathbf{k}}$ denotes the number of $\mathbf{k}$
points used for the irreducible Brillouin zone integration. The
calculated real part of electron susceptibility is shown in
Fig.~\ref{fig4}, which demonstrates that the peak of electron
susceptibility is suppressed as increase the Pd dopant
concentrations. This result along with our experimental facts
suggest the electron susceptibility of system originating from the
itinerant electrons with strong spin-orbit coupling, which is
responsible for superconductivity, is weakened by introducing the
electron correlation, and thus suggesting this system displays a
competition between the electron correlation and spin-orbit
coupling, which seems to be unfavorable for superconductivity.

\section{Conclusion}

In conclusion, we have successfully substituted Pd elements into the
position of Pt in the strong spin-orbit coupling superconductor
SrPt$_3$P, and found that the doping of Pd not only leads to the
change of lattice parameters, but also suppresses the
superconducting transition temperature $T_c$. In addition, the band
structure calculations reveal that the calculated strength of
electron susceptibility is suppressed as increase the Pd dopant
concentrations. These results suggest that the competition between
spin-orbit coupling and electron correlation plays a vital role in
superconductivity of the present $5d$ electron system.

\begin{acknowledgments}
This work is supported by the Natural Science Foundation of China
(No. 11204338, 11227902, and 11404359), the ``Strategic Priority
Research Program (B)" of the Chinese Academy of Sciences (No.
XDB04040300), and the Youth Innovation Promotion Association of the
Chinese Academy of Sciences (No. 2015187). This work is partly
sponsored by the Science and Technology Commission of Shanghai
Municipality (No. 14DZ2260700 and 14521102800).

\end{acknowledgments}

\end{document}